\documentclass[aps,prb,twocolumn,10pt,superscriptaddress,amsmath,amssymb,longbibliography]{revtex4-2}

\usepackage[hidelinks]{hyperref}
\usepackage[utf8]{inputenc}
\usepackage{amsmath}
\usepackage{physics}
\usepackage{xspace}
\usepackage{graphicx}
\usepackage{bbold}
\usepackage{enumitem}
\usepackage{booktabs}
\usepackage{xcolor}

\newcommand{\im}{\ensuremath{\mathrm{i}}}

\DeclareMathOperator{\arcsinh}{arcsinh}

\begin{document}

\title{Magnetic $2\pi$ domain walls for tunable Majorana devices}
\date{\today}
\author{Daniel Hauck}
\affiliation{Institute of Theoretical Solid State Physics, Karlsruhe Institute of Technology, 76131 Karlsruhe, Germany}

\author{Stefan Rex}
\affiliation{Institute of Theoretical Condensed Matter Physics, Karlsruhe Institute of Technology, 76131 Karlsruhe, Germany}
\affiliation{Institute for Quantum Materials and Technology, Karlsruhe Institute of Technology, 76131 Karlsruhe, Germany}

\author{Markus Garst}
\affiliation{Institute of Theoretical Solid State Physics, Karlsruhe Institute of Technology, 76131 Karlsruhe, Germany}
\affiliation{Institute for Quantum Materials and Technology, Karlsruhe Institute of Technology, 76131 Karlsruhe, Germany}

\begin{abstract}
 Identifying realistic platforms capable of controlled operations with Majorana bound states is a key challenge in the study of topological superconductivity. Among the most promising proposals are magnet-superconductor hybrid devices, which employ  magnetic textures to engineer regions of non-trivial topology. Here, we consider the remarkably simple case of $2\pi$ domain walls in a magnetic ribbon placed on a superconducting substrate. We show that for properly chosen parameters, such domain walls generate topological quasi-one dimensional superconducting wires and give rise to localized Majorana bound states at the ribbon edges. Magnetic $2\pi$ domain walls are easily created and controlled with existing experimental techniques, thus providing a versatile platform for Majorana manipulations.
\end{abstract}

\maketitle

\section{Introduction}

Generating topological superconductivity \cite{Qi:2010qag} and Majorana modes \cite{Kitaev:2000nmw, Ivanov:2000mjr} for new quantum technologies, i.e. non-Abelian quantum computation in particular \cite{Nayak:2008zza}, remains a topic of profound interest.
It has inspired numerous proposals for topological nano-heterostructures, including magnet-superconductor hybrid devices. 
In the simplest case, homogeneous ferromagnetic islands are sufficient to induce a topological phase in a superconducting substrate \cite{Rontynen:2015, Li:2016}. One-dimensional chiral Majorana modes are then expected at the rims of the island, and corresponding signatures have indeed been reported by experiments \cite{Menard:2016ihv, Palacio-Morales:2019}. More involved but particularly interesting are magnetic textures consisting of smooth non-collinear arrangements of the magnetization. The continuous 
rotation of the magnetic moments contributes here to the effective spin-orbit coupling  \cite{Braunecker:2010, Choy:2011, Kjaergaard:2011yf} as well as the chemical potential. Depending on the orientation of the local magnetic order, this enables the spatial separation of the superconductor into topologically trivial and non-trivial  regions.
This has been theoretically shown for several extended periodic textures, including Skyrmion lattices \cite{Mascot:2021}, helical and cycloidal order \cite{Rex:2020utd}, and triple-$q$ materials \cite{Bedow:2020}. In all these cases, topological regions are encircled by a number of chiral Majorana modes corresponding to the Chern integer \cite{Qi:2010qag} of the phase.

Intriguingly, a suitably textured magnet-super\-conductor hybrid can also generate spatially localized Majorana bound states (MBSs). On the one hand, the core of a Skyrmion \cite{Yang:2016ltv,Garnier:2019} or a Skyrmion-vortex pair \cite{Hals:2016jkl,Baumard:2019,Rex:2019,Garnier:2019a,Nothhelfer:2021nqb}
can serve as a point-like defect hosting an MBS with its partner state delocalized at the rim of the Skyrmion. Skyrmion-vortex pairs have recently been observed \cite{Petrovic:2021a}, and alternative materials are under investigation \cite{Herve:2018,Kubetzka:2020}. On the other hand, elongated Skyrmions \cite{Gungordu:2017atn} or stripes in a helical phase \cite{Rex:2020utd} can form quasi-one-dimensional topological sub-systems with MBS at their terminations or possibly junctions. Such emergent wires resemble the original nanowire models \cite{Lutchyn:2010xey,Oreg:2010xfn} on an effective level.
Magnetic adatom chains \cite{Peng:2015,Pawlak:2019fhb,Awoga:2023gnv,Kreisel:2021} provide a related platform which uses the same principle.
These setups have also been realized in experiments \cite{Schneider:2020a,Schneider:2021wei} so far, however, without evidence for MBSs.

\begin{figure}
    \centering
    \includegraphics[width=\linewidth]{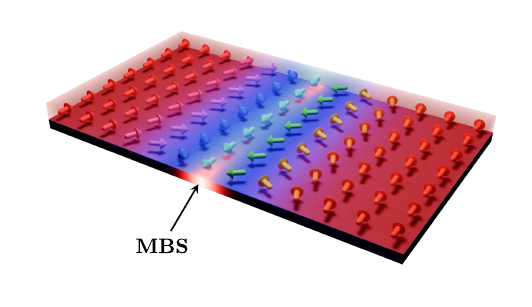}
    \caption{Magnetic-superconducting hybrid heterostructure consisting of a ferromagnetic ribbon on top of a superconducting film. A magnetic $2\pi$ domain wall induces a quasi-one-dimensional topological superconducting region, resulting in spatially localized Majorana bound states (MBSs) at the two ribbon edges. }
    \label{fig:intro}
\end{figure}

In the present work, we propose a relatively simple setup of a magnetic texture --  a $2\pi$ magnetic domain wall -- for the generation of such an effective wire realization hosting 
localized MBS, see Fig.~\ref{fig:intro}.
Specifically, we consider a ferromagnetic ribbon hosting a $2\pi$ domain wall on a superconductor and identify suitable conditions for the generation of MBS at the ribbon edge, where the domain walls terminate.  The $2\pi$ domain wall textures have clear advantages for the generation of MBS: $(i)$ they are readily available and do not require a special design of the magnetic interactions, $(ii)$ they can be easily driven by spin-currents \cite{Ryu2013,Emori2013} thus enabling, in the adiabatic limit, a dynamic manipulation of the MBS's positions. The latter will be eventually important for the implementation of braiding operations.

The remainder of this paper is organized as follows. We introduce the proposed setup and specify the model as well as its parameters in sections \ref{sec:motivation} and \ref{sec:Setup}.
In section \ref{subsec:1dDoubleIsingWall} we show that the $2\pi$ domain wall generates an effective quasi-one dimensional wire. We discuss its phase diagram and determine the parameter regime where it is topologically non-trivial, thus leading to the emergence of MBSs. A full numerical solution of the finite size system is presented in section \ref{sec:results}, corroborating our findings.
Finally, section \ref{sec:conclusion} closes with a summary and a discussion of perspectives for Majorana operation exploiting the manipulation of $2\pi$ domain walls with existing experimental techniques.

\section{Magnet-superconductor heterostructure}

\subsection{Motivation}
\label{sec:motivation}

Our starting point is the standard Bogoliubov-de-Gennes Hamiltonian for a two-dimensional superconductor in the presence of a Rashba spin-orbit coupling $\alpha$ and an effective magnetic field $\vec h(\vec r)$,
\begin{align}
    &H  =  \label{eq:ham}
    \\ \nonumber 
    & \Big(-\frac{\nabla^2}{2m}-\mu\Big)\tau_z +\Delta\tau_x 
    -\im\alpha  (\sigma_x \partial_y - \sigma_y \partial_x) \tau_z 
    + \vec h(\vec r)\, \vec \sigma.
\end{align}
The Hamiltonian acts on the spinor $(c_\uparrow, c_\downarrow, c^\dagger_\downarrow, - c^\dagger_\uparrow)$ defined in terms of annihilation, $c_{\sigma}$, and creation operators, $c_\sigma^\dagger$, with $\sigma \in \{\uparrow, \downarrow\}$. Here, $m$ is the effective mass of the electron, $\mu$ is the chemical potential, and $\Delta$ is the pairing amplitude that has been chosen to be real. The matrices in spin and particle-hole space, respectively, are given by 
\begin{align}
    \sigma_i & = \mathbb{1}_2\otimes\Sigma_i,\quad 
    \tau_i    = \Sigma_i\otimes\mathbb{1}_2
\end{align}
with the standard $2\times 2$ Pauli matrices $\Sigma$ and the Kronecker product $\otimes$.

\begin{figure}
    \centering
        \includegraphics[width=\linewidth]{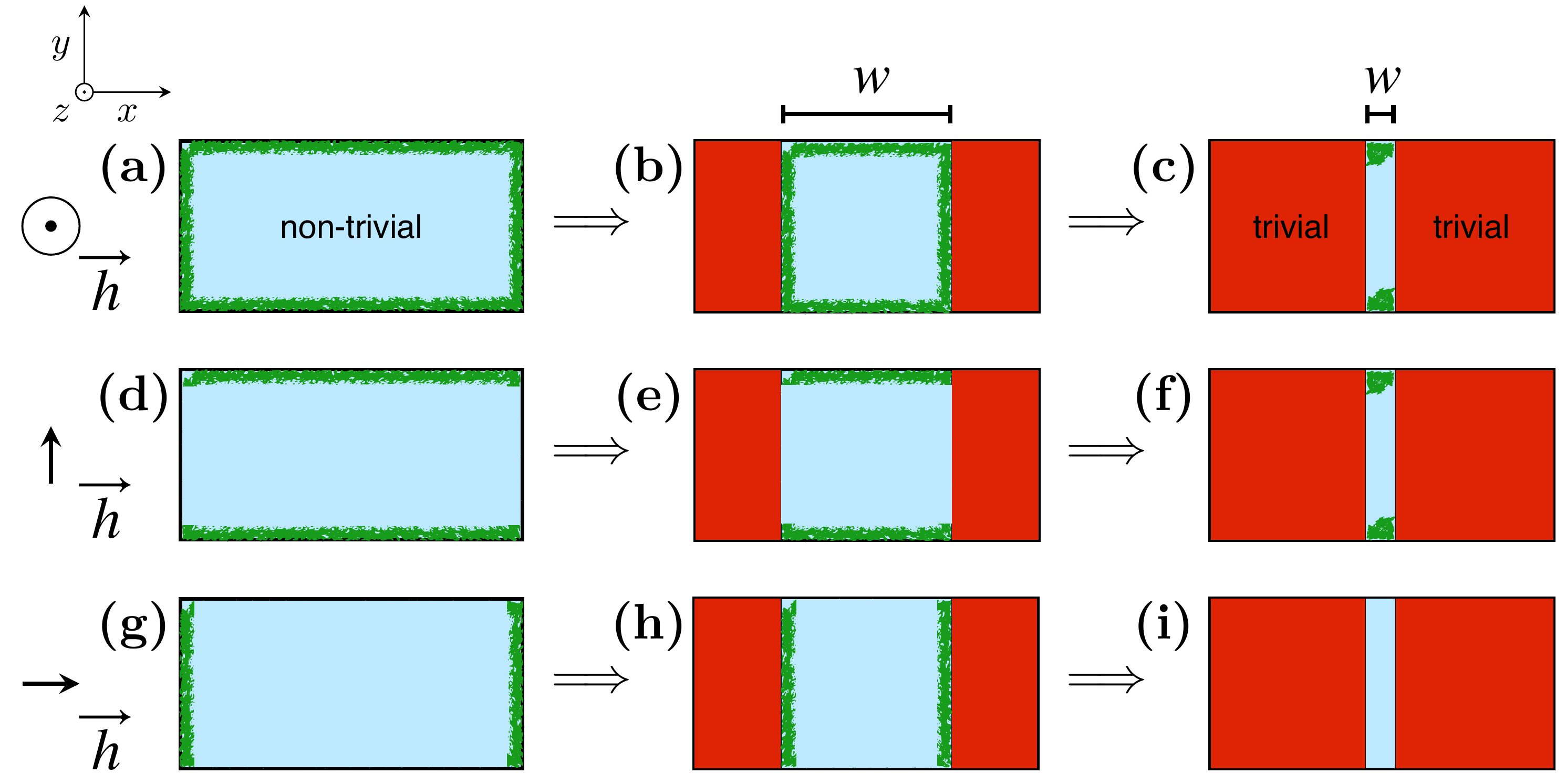}
    \caption{Illustration of generating MBSs by reducing the region of a topologically non-trivial superconductor within a ribbon. The blue and red regions indicate topologically non-trivial and trivial superconducting regions, respectively. The green shading represents the presence of Majorana edge modes. By shrinking the topologically non-trivial region with the help of a spatially dependent out-of-plane field (a)-(c) or in-plane field perpendicular to the ribbon (d)-(f) MBSs are realized. This is not the case for a spatially dependent in-plane field pointing along the ribbon (g)-(i). The spatial dependence of the effective field can be realized with the help of the exchange field generated by a $2\pi$ magnetic domain wall, see Fig.~\ref{fig:intro}. 
    } 
    \label{fig2}
\end{figure}

In order to set the stage and to illustrate the main idea, consider first the situation of a constant homogeneous field $\vec h(\vec r) \equiv \vec h$. In the absence of the Rashba coupling, $\alpha = 0$, the lowest positive eigenenergy is given by 
$| \sqrt{\Delta^2 + (\vec k^2/(2m) - \mu)^2} - |\vec h| |$. Of interest to us here is the behavior of this eigenenergy at zero wavevector $\vec k = 0$. It vanishes at a critical field with absolute value
\begin{align} \label{eq:hcr}
h_{cr} \equiv \sqrt{\Delta^2 + \mu^2},
\end{align}
irrespective of the orientation of $\vec h$. In the presence of a finite Rashba coupling $\alpha$, this vanishing of the eigenenergy at $h_{cr}$ becomes a transition between a topologically trivial and non-trivial superconductor for $|\vec h| < h_{cr}$ and $|\vec h| > h_{cr}$, respectively. 

The case of an out-of-plane field $\vec h = h_z \hat e_z$ was discussed by Alicea \cite{Alicea:2009dvr}. 
The Hamiltonian is then equivalent to two spinless $p_x+i p_y$ superconductors \cite{bookShen} where one of them might be tuned to a topologically non-trivial phase, $h_z > h_{cr}$, where a delocalized Majorana mode at the edge of the system will be present. We will assume here that the system is given by a ribbon aligned along, say, the $x$-direction, see Fig.~\ref{fig2}(a) where the green shaded edge represents the Majorana edge mode. The topological superconductor can be reduced to a region with finite width $w$, see Fig.~\ref{fig2}(b), by sandwiching it between two regions with a lower field value $h_z < h_{cr}$, which might be achieved with an inhomogeneous field configuration $h_z(x)$. If this region shrinks further such that $w$ becomes on the order of the Majorana localization length along the ribbon, i.e., the $x$-direction, the Majorana edge modes hybridize \cite{Potter2010}. We thus obtain an effective quasi-one dimensional, topologically non-trivial superconducting wire that possesses localized MBSs at its termination points, see Fig.~\ref{fig2}(c). We propose that the spatial profile for $h_z(x)$ required for this scenario can be realized with the help of the exchange field generated by a $2\pi$ domain wall. 

However, out-of-plane fields have the disadvantage that they might lead to pair-breaking in the superconductor due to orbital-limiting effects \cite{Loder:2015}. For this reason, we focus in the following on in-plane fields $\vec h \perp \hat e_z$ for which the topological properties have been also discussed in the literature before, and we refer the reader to Refs.~\cite{Loder:2015,Deng:2016,Yuan:2018}. In this case, the superconductor is fully gapped only for small fields, $|\vec h| < h_{cr}$. For fields exceeding the critical field, $|\vec h| > h_{cr}$, the superconductor is only partially gapped, resulting in a Bogoliubov Fermi surface. For wavevectors along the in-plane field $\vec h$ the superconductor possesses a gap, but this gap closes when turning the wavevector away from $\vec h$ resulting in a Fermi arc shaped surface. Nevertheless, the superconductor still possesses a weak non-trivial topology \cite{Fu2007}. If a constant field is applied perpendicular to the ribbon, say, in $y$-direction, $\vec h = h_y \hat e_y$, with $h_y > h_{cr}$, a Majorana edge state is induced but only along the upper and lower edge of the ribbon, see Fig.~\ref{fig2}(d). With the help of a spatially dependent $h_y(x)$ the extent of the topologically non-trivial region can be reduced as before until we end up with localized MBSs, see Fig.~\ref{fig2}(e) and (f). This cannot be achieved for a field pointing along the $x$-direction, i.e., along the ribbon because in this case, the Majorana modes would hybridize away upon shrinking, see Fig.~\ref{fig2}(g)-(i). For this reason, we concentrate in the following on effective fields $\vec h$ that are mostly aligned perpendicular to the ribbon.

The configuration with a constant in-plane field $\vec h = h_y \hat e_y$ realizes a weak topological insulator. For a vanishing wavevector $k_x = 0$ along the ribbon, the Hamiltonian in Fourier space then reduces to 
\begin{align}  \label{eq:ham2}
    H_{\perp}(k_y)  = 
    \Big(\frac{k_y^2}{2m}-\mu\Big)\tau_z +\Delta\tau_x 
     + \alpha \sigma_x k_y \tau_z 
    + h_y \sigma_y.
\end{align}
This Hamiltonian possesses particle-hole symmetry $\tau_y \sigma_y H_\perp(k_y) \tau_y \sigma_y = - H^*_\perp(-k_y)$ and an effective time-reversal symmetry $H^*_\perp(-k_y) = \sigma_z H_\perp(k_y) \sigma_z$. The latter arises as the complex conjugation and the reversal of $k_y \to -k_y$ can be compensated by a rotation of spin space by $\pi$ around the $z$-axis. (This might be more easily seen by making the Hamiltonian real by applying a $\pi/2$-rotation in spin space $\sigma_x \to \sigma_y$ and $\sigma_y \to - \sigma_x$.) Finally, we have the chiral symmetry $\tau_y \sigma_x H_\perp(k_y) \tau_y \sigma_x = -  H_\perp(k_y)$. This puts the Hamiltonian in the BDI class of the band topology classification \cite{bookShen} and its topology can be characterized by an integer winding number. This is at the origin of the appearance of Majorana modes perpendicular to the ribbon, see Fig.~\ref{fig2}(e).

\subsection{Setup of the magnet-superconducting heterostructure}
\label{sec:Setup}

\subsubsection{Exchange field of a $2\pi$ domain wall}

\begin{figure}
    \begin{center}
    \includegraphics[width=0.9\linewidth]{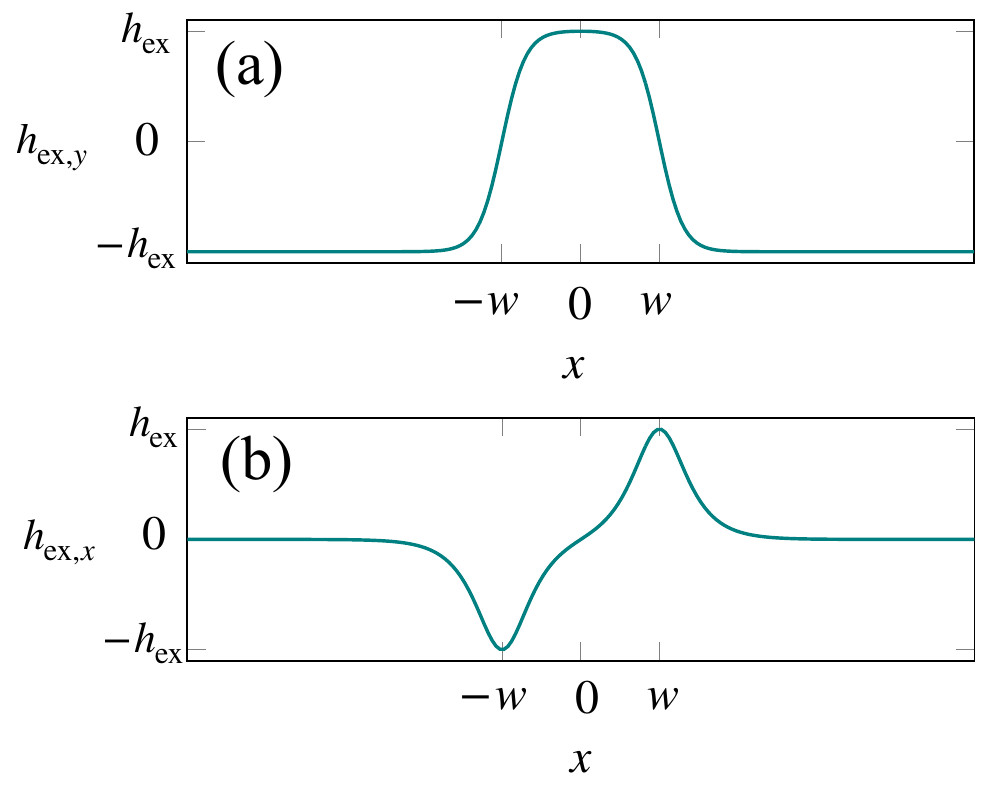}
    \end{center}
    \caption{Spatial dependence of the exchange field of Eq.~\eqref{eq:mag} due to the presence of a $2\pi$ domain wall for the ratio $x_0/w = 0.1$. (a) $y$-component and (b) $x$-component of $\vec h_{\rm ex}(x)$. Away from the domain wall the exchange field reduces the total field \eqref{eq:totfield} whereas at its center it is enhanced. The $x$-component is only finite close to the center of the two individual $\pi$ domain walls located at $\pm w$. }
    \label{figWall}
\end{figure}

In order to generate an effective field $\vec h(x)$ that depends on the $x$-coordinate in such a manner that it 
gives rise to the scenario depicted in Fig.~\ref{fig2}(f), we assume that it is composed of an external field $H$ applied along the $y$-direction and an exchange field, 
\begin{align} \label{eq:totfield}
\vec h(x) = h_{0} \hat e_y + \vec h_{\rm ex}(x), 
\end{align}
with $h_{0} = \frac{g}{2} \mu_B \mu_0 H > 0$ where $g$ is the g-factor, $\mu_B$ is the Bohr magneton, $\mu_0$ is the magnetic constant. As the exchange field $\vec h_{\rm ex}$ should reduce the total field at the outer regions for $|x| \to \infty$, it should be anti-aligned with the applied field $h_0$. This can be achieved with the help of an antiferromagnetic exchange coupling between the superconductor and the magnet of the heterostructure, i.e., $\vec h_{\rm ex} = - J \vec M(x)$ with $J > 0$ and the magnetization $\vec M(x)$. In the ground state the magnetization is aligned with $h_0$, thus reducing the total field $|\vec h|$. In the following, we treat the magnetization in the mean-field approximation and neglect any fluctuations of $\vec M$.

We assume a magnet with a strong easy-plane anisotropy, such that the magnetization is confined within the plane. Moreover, the magnet should contain a $2\pi$ domain wall positioned at $x=0$, whose exchange field reads 
\begin{align}
\vec h_{\rm ex}(x) &= - h_{\rm ex}
\begin{pmatrix}
        \sin\theta(x) \\
        \cos\theta(x) \\
        0
    \end{pmatrix} ,
    \label{eq:mag}
\end{align}
with $h_{\rm ex} = J M_s > 0$, where $M_s$ is the saturation magnetization, and the spatial $x$-dependence of the angle is parametrized by 
\begin{align}
    \theta   & = 2\arctan\left(\frac{\sinh \frac{x}{x_0} }{\sinh \frac{w}{x_0}} \right) - \pi .
        \label{eq:theta}
\end{align}
This describes a pair of $\pi$ Bloch domain walls separated by the distance $w$, and whose individual width is characterized by $x_0$. 
The angle $\theta$ increases from $-2\pi$ to $0$ for increasing coordinate $x$ so that the exchange field points along the negative $y$-direction far away from the wall, $\vec h_{\rm ex}(x) \to - h_{\rm ex} \hat e_y$. At the center $\theta(0) = -\pi$ and $\vec h_{\rm ex}(0) = h_{\rm ex} \hat e_y$ points along the positive $y$-direction, thus enhancing the total field. For an illustration, see Fig.~\ref{figWall}.

The profile \eqref{eq:theta} corresponds to a solution of the double sine-Gordon model, and its parameters, $w$ and $x_0$, can be expressed in terms of micromagnetic constants and the applied field, see Ref.~\cite{Muller:2016}. For a ferromagnet described by the exchange stiffness $A$, easy-axis anisotropy $K_y$ within the $x$-$y$ plane, and a field $H$ applied along the $y$-direction the domain wall parameters are given by \cite{Muratov:2008,Muller:2016},
\begin{align}
w &= \sqrt{\frac{2 A}{2K_y + \mu_0 H M_s}} \arcsinh\sqrt{ \frac{2K_y + \mu_0 H M_s}{\mu_0 H M_s} },
\\
x_0 &= \sqrt{\frac{2 A}{2K_y + \mu_0 H M_s}}.
\end{align}

\subsubsection{Boundary condition on the ribbon edges }

\begin{figure}
    \begin{center}
    \includegraphics[width=\linewidth]{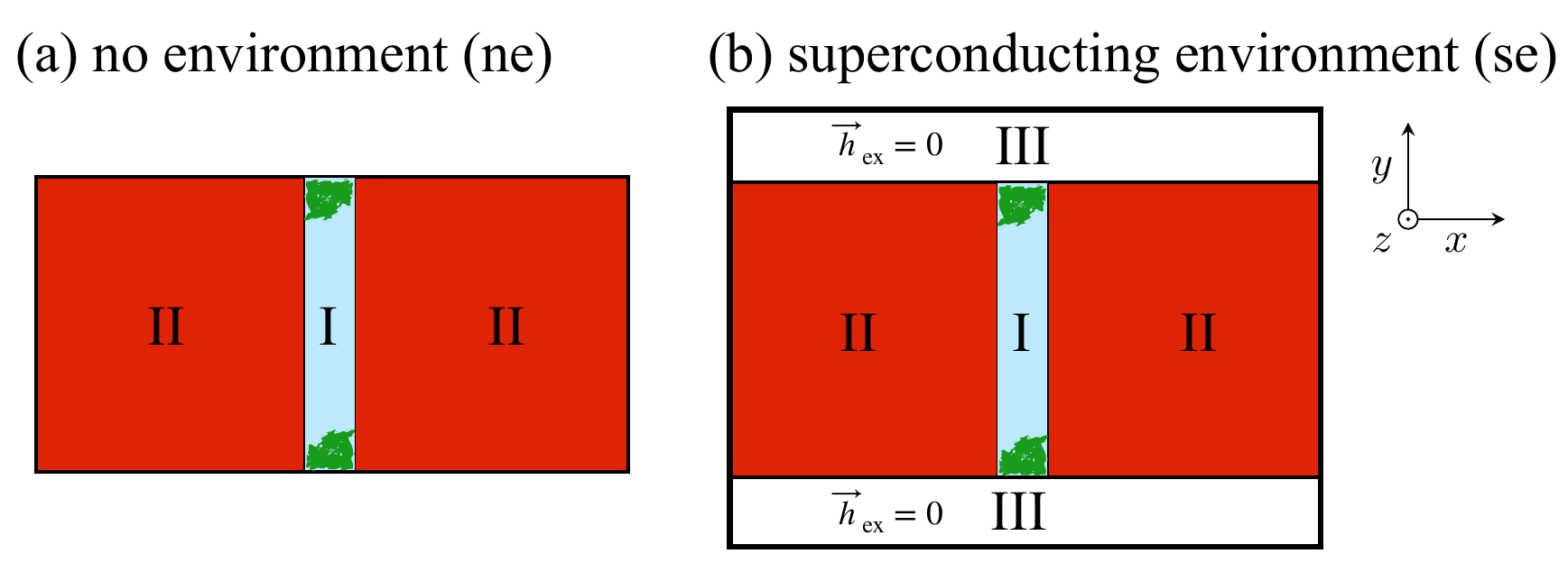}
    \end{center}
    \caption{Two boundary conditions for the heterostructure of Fig.~\ref{fig2}(f): (a) the ribbon is surrounded by vacuum, corresponding to no environment (ne) and (b) the extension of the superconducting substrate is larger than the magnetic ribbon leading to a superconducting environment where $\vec h_{\rm ex} = 0$. 
    At the three indicated positions the field \eqref{eq:totfield} is pointing in $y$-direction with components $h_{{\rm I},y} = h_0 + h_{\rm ex}$, $h_{{\rm II},y} = h_0- h_{\rm ex}$, and $h_{{\rm III},y} = h_0$.}
    \label{figBC}
\end{figure}

For the ribbon edges, we consider two types of boundary conditions. For the first type, we assume that the hybrid structure housing the domain wall is surrounded by vacuum corresponding to no environment (ne), see Fig.~\ref{figBC}(a). For the second type, we assume that the superconducting substrate extends beyond the width of the magnetic ribbon, see Fig.~\ref{figBC}(b), resulting in a superconducting environment (se). Within the superconducting environment the exchange field vanishes leading to an effective field just given by the externally applied field, $\vec h(\vec r) = h_{0} \hat e_y$ for $\vec r$ within (se). 

\subsubsection{Necessary conditions for the realization of MBSs}

Far away from the domain wall in region II as well as in region III within the superconducting environment for (se) boundary conditions, see  Fig.~\ref{figBC}, the field $\vec h$ is basically constant and pointing in $y$-direction with values  $h_{{\rm II},y} = h_0 - h_{\rm ex}$ and $h_{{\rm III},y} = h_0$, respectively. A prerequisite for the appearance of MBSs is that both regions are in a topologically trivial phase and, consequently, their fields are below the critical field, 
\begin{align} \label{eq:conditions}
|h_{{\rm II},y}| < h_{cr}, \quad h_{{\rm III},y} < h_{cr}.
\end{align}
The latter avoids that the Majorana mode is spilling into the superconducting environment.
This poses necessary conditions that must be fulfilled for the realization of MBSs in our proposed setup.

In the limiting case $w \gg x_0 $ of two sharp $\pi$ domain walls, an inner region I emerges where the field is also approximately constant and given by $h_{{\rm I},y} = h_0 + h_{\rm ex}$. If the width $w$ exceeds any relevant coherence length of the superconductor, we might expect MBSs if the field exceeds the critical field such that region II is in the topologically non-trivial phase,
\begin{align} \label{eq:condition2}
h_{{\rm I},y} > h_{cr} .
\end{align}
This also serves as a bound for the appearance of MBSs. 

\begin{figure}
    \begin{center}
    \includegraphics[width=\linewidth]{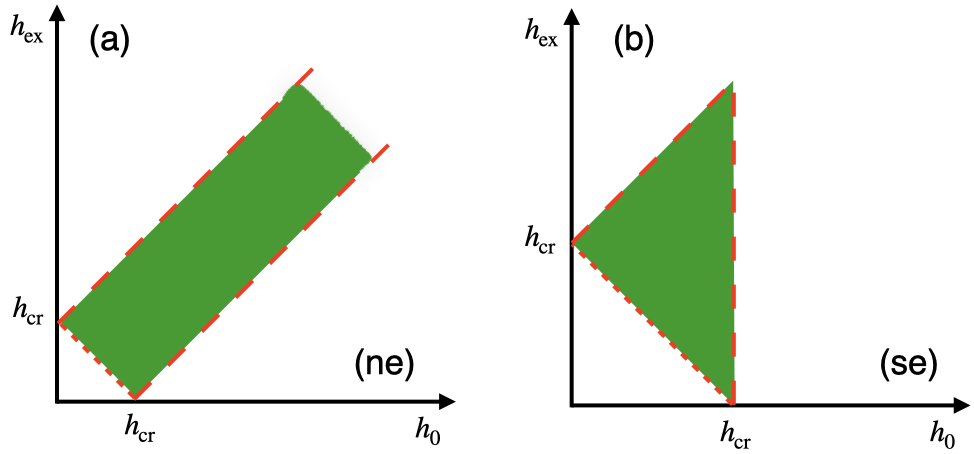}
    \end{center}
    \caption{Necessary conditions for the appearance of MBSs are fulfilled in the region enclosed by the red 
   dashed lines corresponding to \eqref{eq:conditions} 
    where 
    $h_0$ is the applied field and $h_{\rm ex}$ is the magnitude of the exchange field. The critical field $h_{cr}$ is given in Eq.~\eqref{eq:hcr}. 
    The bound \eqref{eq:condition2} shown by the dotted line delimits the region from below, thus yielding the green shaded parameter regime. 
       The superconducting environment (se) yields additional constraints, thus disfavouring MBSs.
}
    \label{figNCOND}
\end{figure}

These conditions are summarized in the diagram shown in Fig.~\ref{figNCOND}. They identify a parameter region within the $(h_0, h_{\rm ex})$ plane for which localized MBSs could be potentially stabilized upon decreasing the width $w$, see Fig.~\ref{fig2}(f). In order to find out for which values of $(h_0, h_{\rm ex})$
MBSs indeed materialize, we will analyze in the next sections the explicit numerical solution of the model in the presence of the $2\pi$ domain wall.

\subsubsection{Numerical implementation and parameters}
\label{Subsec:Parameters}

In the following, we will use units of energy and length where $\Delta = 1$ and the mass $m = 1$ (with reduced Planck constant $\hbar = 1$). Moreover, we will limit ourselves to a chemical potential $\mu = 0$ \cite{Alicea:2009dvr}, and for the Rashba coupling we will use $\alpha = 0.3$ and sometimes compare with $\alpha = 0$. For the numerical implementation, we discretized the Hamiltonian in a standard fashion using two-point stencils on a regular two-dimensional grid with lattice constant $a$ promoting the model to a tight-binding Hamiltonian, that coincides with Eq.~\eqref{eq:ham} in the continuum limit.
The software and data used for this work are available in Ref.~\cite{dataupload}.
This leaves us with three parameters: the lattice constant $a$ and two length scales characterizing the domain wall, the width $w$ and the steepness $x_0$. We will consider three distinct regimes,
\begin{align} \label{Eq:WallTypes}
\begin{array}{ll}
w > a \gg x_0 & \qquad \text{double Ising wall},\\
w > x_0 \gg a & \qquad \text{wide $2\pi$-wall},\\
w \sim x_0 \gg a & \qquad \text{narrow $2\pi$-wall}. 
\end{array}
\end{align}
In the first case, the winding happens basically abrupt from one lattice site to the next such that the magnet is rather an Ising magnet and the wall structure corresponds to two consecutive Ising walls. This limit is numerically the best accessible, and we will present most results for this case. The second and third cases rather correspond to the continuum Hamiltonian of Eq.~\eqref{eq:ham} and describe proper $2\pi$ domain walls that are wide or narrow depending on the ratio of the two domain wall parameters, $w$ and $x_0$.

\subsection{Effective quasi-one dimensional superconducting wire}
\label{sec:Eff1dWire}

The $2\pi$ domain wall breaks the translational symmetry of the system along the ribbon, i.e., the $x$-axis. It is expected that the wall results in an effective potential that might possess a bound state, thus leading to in-gap states localized along the $x$-direction but delocalized perpendicular to the ribbon, i.e., along the $y$-axis. Such an in-gap state thus gives rise to an effective  low-energy theory that can be identified with a quasi-one dimensional superconducting wire. 

For its derivation, we consider a partial Fourier transform of the Bogoliubov-de-Gennes Hamiltonian \eqref{eq:ham} with respect to the $y$-coordinate,
\begin{align}\label{eq:hamFT}
    H(k_y)  = & \Big(\frac{- \partial_x^2 + k_y^2 }{2m}-\mu\Big)\tau_z 
    \\ \nonumber 
    & +\Delta\tau_x 
    + \alpha  (\sigma_x k_y + \im \sigma_y \partial_x) \tau_z 
    + \vec h(x)\, \vec \sigma.
\end{align}
The corresponding eigenvalue problem, $H(k_y) \vec \phi_{s n k_y}(x) = \varepsilon_{s n k_y} \vec \phi_{s n k_y}(x)$, with the field $\vec h(x)$ given by Eq.~\eqref{eq:totfield} can be solved numerically. Eigenvalues and -states are labeled with $k_y$ and a band index $n = 1,2,3,...$. 
The Hamiltonian possesses the standard particle-hole symmetry
 $\tau_y \sigma_y H(k_y) \tau_y \sigma_y = - H^*(-k_y)$.
 As a consequence, 
eigenvalues come in pairs $\varepsilon_{s n k_y} = s \varepsilon_{n k_y}$ where $s = \pm 1$. The corresponding eigenfunctions are related via $\vec \phi_{- n k_y} = \tau_y \sigma_y \vec \phi^*_{+ n k_y}$;
for later convenience,  we explicitly note  that $\tau_y \sigma_y  (\phi^*_1, \phi^*_2,\phi^*_3,\phi^*_4)^T = (-\phi^*_4, \phi^*_3,\phi^*_2,-\phi^*_1)^T$.

\begin{figure}
    \centering
        \includegraphics[width=\linewidth]{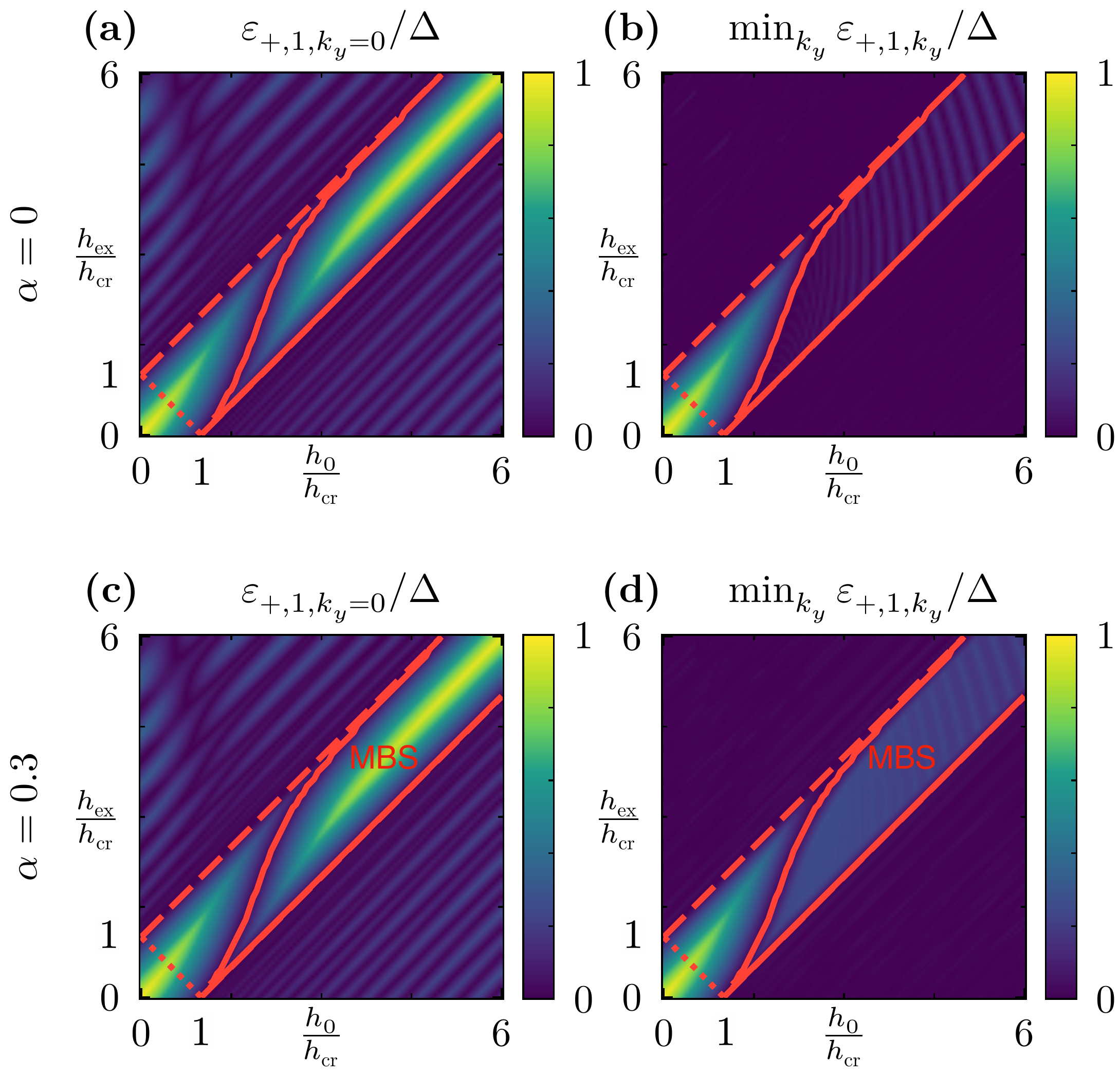}

    \caption{Lowest positive eigenenergy $\varepsilon_{+,1,k_y}$ at $k_y = 0$, (a) and (c) and at its minimum, (b) and (d),
    as a function of applied field $h_0$ and exchange field $h_{\rm ex}$ for a parameter set of a double Ising wall and two values of the Rashba spin-orbit coupling $\alpha = 0$, upper row, and $\alpha = 0.3$, lower row. The density plots show the value of the energy in units of $\Delta$, as indicated by the color bar on the right-hand side. For $|h_0 - h_{\rm ex}| > h_{cr}$ the superconductor is gapless and the energy corresponds to a scattering state. 
For $|h_0 - h_{\rm ex}| < h_{cr}$ the superconductor is gapped, and we find in-gap eigenstates that are bound to the domain wall. Its eigenenergy vanishes at the red solid line, and a robust topologically non-trivial superconducting wire with MBSs emerges  for finite $\alpha$.
}
    \label{fig6}
\end{figure}

\subsubsection{Quasi 1d superconducting wire along a double Ising wall}
\label{subsec:1dDoubleIsingWall}

In the following, we present the numerical solution of the eigenvalue problem obtained after discretizing the system along the $x$-direction. We use the parameters specified in section \ref{Subsec:Parameters} and start with a discussion of the double Ising wall limit. 
Discretizing the $x$-direction with lattice constant $a=0.1$ and $N_x = 200$ lattice points amounting to a system size $a N_x = 20$, we consider a double Ising wall parametrized with a steepness $x_0 = 0.02 \ll a$ and a width $w = 0.2$. The distance between the two Ising walls is on the order of the lattice constant such that the total wall corresponds basically to a local scattering potential. 

We focus on the eigenstate with the lowest positive energy, $\varepsilon_{+,1,k_y}$, that eventually determines the effective low-energy theory. In Fig.~\ref{fig6} we show the evolution of both the minimum of $\varepsilon_{+,1,k_y}$ and its value at $k_y=0$ within the ($h_{\rm ex}$,$h_0$) plane and compare two values for the Rashba spin-orbit coupling, $\alpha = 0$ for the panels in the upper row and $\alpha = 0.3$ for the panels in the lower row. 
In the outer region $|h_0 - h_{\rm ex}| > h_{cr}$ the superconductor is gapless and there exists a value for $k_y$ where the energy $\varepsilon_{+,1,k_y}$ vanishes, see Fig.~\ref{fig6}(b) and (d). In the central region $|h_0 - h_{\rm ex}| < h_{cr}$ enclosed by the dashed lines, the superconductor is gapped, and only here we find an in-gap eigenstate that is localized at the domain wall. The spectrum at $k_y = 0$ looks quite similar for the cases with and without Rashba spin-orbit coupling, see Fig.~\ref{fig6}(a) and (c), but only in the former case a topologically non-trivial situation can be expected for the double Ising wall. Importantly, the eigenvalue $\varepsilon_{+,1,k_y=0}$ vanishes at the red solid line within the central region $|h_0 - h_{\rm ex}| < h_{cr}$ such that for larger $h_0 = h_{\rm ex}$, one enters a regime where the necessary conditions \eqref{eq:conditions} and \eqref{eq:condition2} for the appearance of MBSs are both  fulfilled. 
This identifies a regime enclosed by the red solid line in Fig.~\ref{fig6}(c) and (d) where MBSs are expected for a finite Rashba coupling $\alpha$. In panel (b) and (d) we show the minimal value of $\varepsilon_{+,1,k_y}$ as a function of $k_y$ and indeed, a robust topological gap is only found for finite $\alpha = 0.3$ in Fig.~\ref{fig6}(d).

Within the central region $|h_0 - h_{\rm ex}| < h_{cr}$ we find an in-gap state that is bound to the domain wall and thus realizes a quasi-one dimensional superconducting wire. We argue that for finite Rashba coupling, this wire undergoes a phase transition at the red solid line between a topologically trivial and non-trivial quasi-one dimensional superconductor. The topologically non-trivial superconductor occupies the upper right region of Fig.~\ref{fig6}(c) and (d) where the necessary conditions of Eqs.~\eqref{eq:conditions} are fulfilled and here MBSs can be expected at the terminations of the quasi-one dimensional wire.

\begin{figure}
    \centering
    \includegraphics[width=\linewidth]{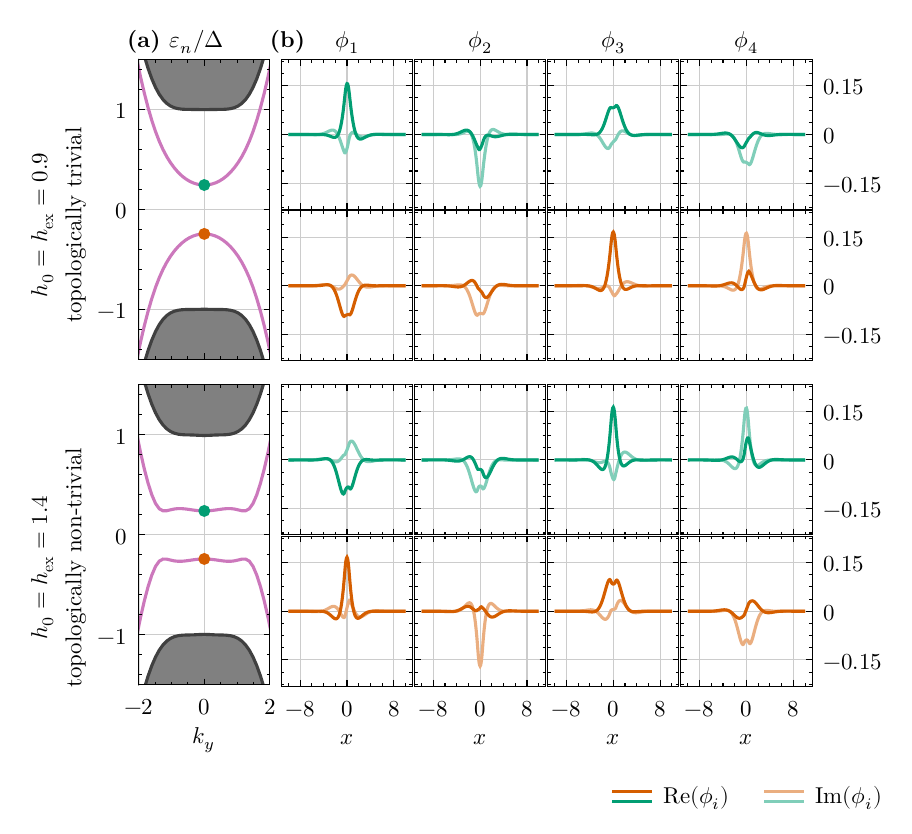}
    \caption{In-gap eigenstates close to the topological phase transition occurring at the red solid line in Fig.~\ref{fig6}(c) and (d).
    The upper and lower panels display the properties within the topologically trivial and non-trivial phases, respectively. The energy spectrum is shown in panel (a) on the left-hand side. The components of the eigenfunctions with lowest eigenenergies are displayed in panel (b) on the right-hand side for $k_y = 0$ as indicated by the dots on the dispersions. Across the transition, the eigenfunctions are swapped between particle-hole partners. Here, the domain width parameter $w = 0.4$.
}
    \label{fig7}
\end{figure}

In order to illustrate the topological phase transition occurring at the red solid line in Fig.~\ref{fig6}(c) and (d) for finite $\alpha$, we discuss the eigenvalues and eigenvectors close to this transition in more detail. Fig.~\ref{fig7}(a) shows the dispersion of the eigenenergies where the pink line represents the in-gap state and its particle-hole counterpart. The upper panel and lower panels correspond to values for $h_0$ and $h_{\rm ex}$ below and above the phase transition. Fig.~\ref{fig7}(b) displays the eigenvector of the in-gap state at $k_y = 0$ as indicated by the dots in (a). The red and orange components of the spinor eigenfunction are consistent with particle-hole symmetry. When crossing the transition, the eigenfunctions remain almost unaffected except that they are swapped between particle-hole partners.

\subsubsection{Quasi 1d superconducting wire along a $2\pi$ domain wall}
\label{subsec:1d2PiWall}

\begin{figure}
    \centering
        \includegraphics[width=\linewidth]{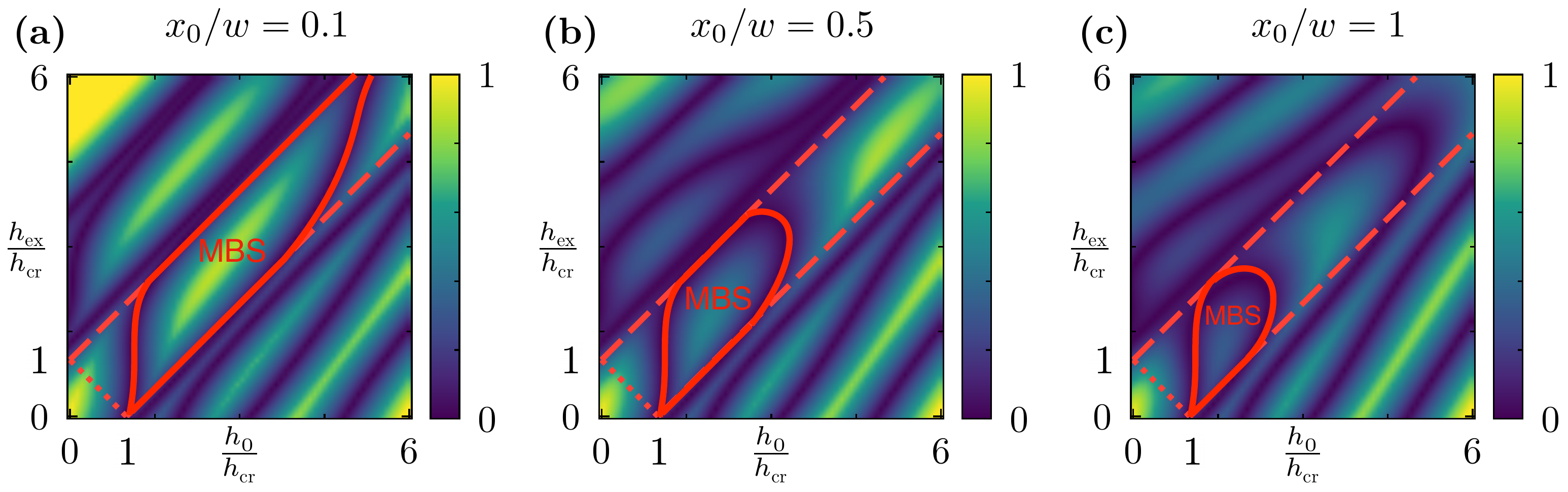}

    \caption{Lowest positive eigenenergy $\varepsilon_{+,1,k_y}$ at $k_y = 0$ for $\alpha = 0.3$ as in Fig.~\ref{fig6}(c) but for a $2\pi$ domain wall. An increasing ratio of the wall parameters $x_0/w$, corresponding to a narrowing of the $2\pi$ domain wall. A state bound to the domain wall is only found within the region 
$|h_0 - h_{\rm ex}| < h_{cr}$ where the superconductor is gapped. The topologically non-trivial regime where MBSs can be expected shrinks as a function of increasing $x_0/w$.
}
    \label{fig1dWire2PiWall}
\end{figure}

We now turn to the situation of a proper $2\pi$ domain wall. Here, we discretize the Hamiltonian \eqref{eq:hamFT} in $x$-direction with lattice constant $a = 0.01$ using $N_x = 500$ lattice points corresponding to a system size $a N_x = 5$. The width of the wall is chosen to be $w = 0.4$ and three values for the steepness are considered $x_0 = 0.04$, $x_0 = 0.2$ and $x_0 = 0.4$ thus illustrating the evolution from a wide to a narrow $2\pi$ domain wall, see Eq.~\eqref{Eq:WallTypes}.

The lowest positive eigenenergy $\varepsilon_{+,1,k_y}$ at $k_y = 0$ for a finite Rashba coupling $\alpha$ is shown in Fig.~\ref{fig1dWire2PiWall} for the three cases, and these results should be compared to Fig.~\ref{fig6}(c) for the double Ising wall.
First of all, it is striking that outside the interesting regime, $|h_0 - h_{\rm ex}| > h_{cr}$, where the superconductor is gapless, the dispersion of the scattering states is more pronounced than before, resulting in a more structured density plot. In the central region $|h_0 - h_{\rm ex}| < h_{cr}$ where the superconductor is gapped, we find again an in-gap state that is bound to the domain wall. 

As a function of increasing $h_0 = h_{\rm ex}$, the in-gap state, however, vanishes again, thus delimiting the candidate region for the emergence of MBSs to higher field values. This region thus shrinks as a function of  increasing $x_0/w$, i.e., when the $2\pi$ domain wall becomes narrower. The in-gap states at higher field values will give rise to effective 1d superconducting wires that might or might not host MBSs. These large field values are likely to be less relevant, but we shortly return to this issue in the next section.

\subsection{Numerical solution of the finite-size magnet-superconducting heterostructure}
\label{sec:results}

\begin{figure}
    \centering
        \includegraphics[width=0.95\linewidth]{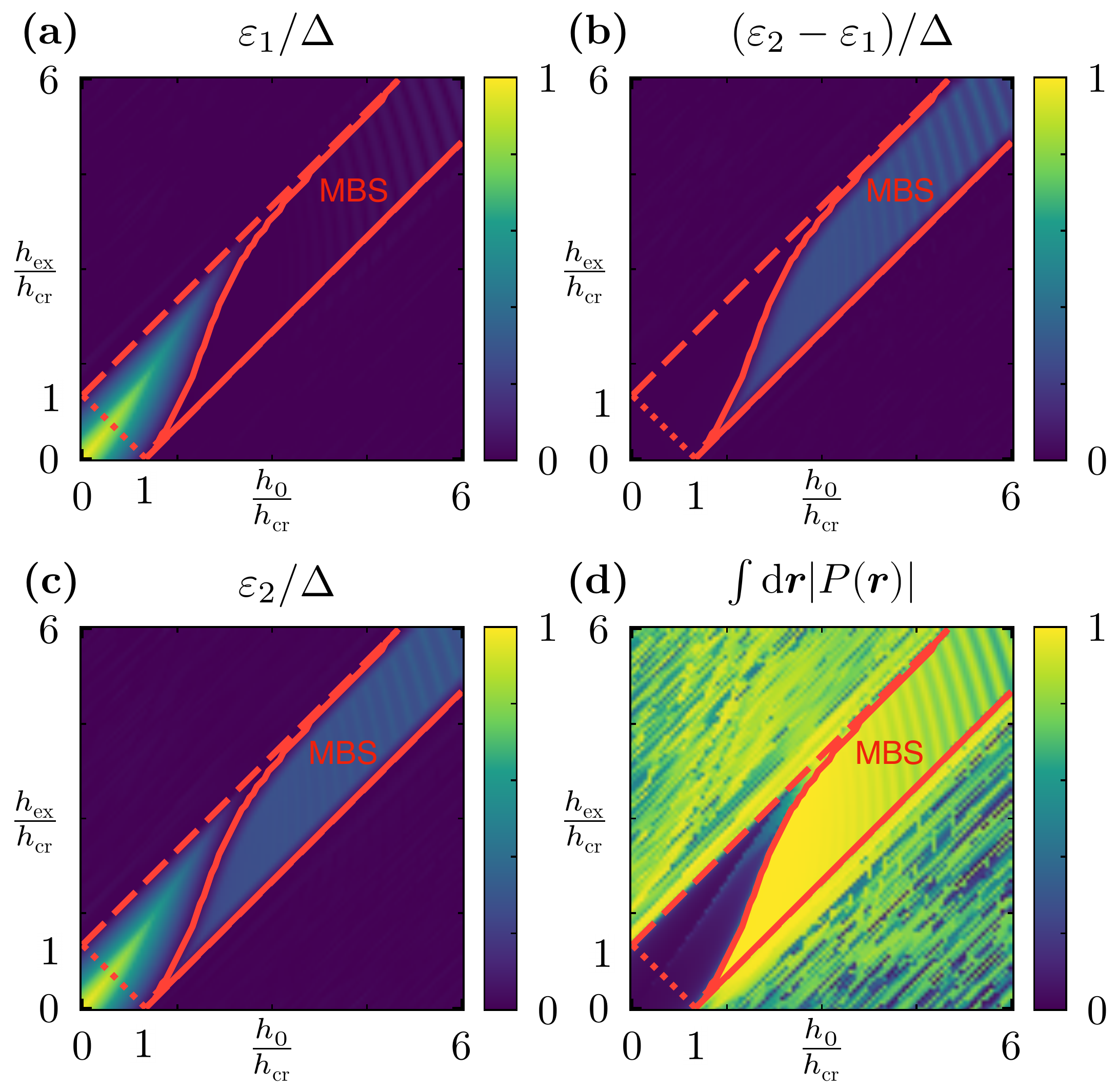}

    \caption{Numerical diagonalization of the finite-size system for a double Ising wall with parameter $w=0.2$. Panel (a)-(c) show the lowest two eigenenergies $\varepsilon_1$ and $\varepsilon_2$ as function of $h_0$ and $h_{\rm ex}$ for a finite Rashba coupling $\alpha = 0.3$.
 A Majorana zero mode $\varepsilon_1 \approx 0$ with a finite topological gap $\varepsilon_2 - \varepsilon_1 > 0$   is only present in the region previously identified to be a candidate for hosting MBS, see Fig.~\ref{fig6}(c). Within this region, the Majorana polarization integrates to unity, see panel (d).
    }\label{fig:Num1}
\end{figure}

We complement our discussion with the results from a full numerical solution of the Hamiltonian \eqref{eq:ham} for a finite system size.
This numerical approach allows confirming and verifing the findings of the previous section as well as to study the impact of the two distinct boundary conditions specified in Fig.~\ref{figBC}. The Hamiltonian is now discretized on a two-dimensional grid with lattice constant $a = 0.1$ and lattice sites $N_x = 200$ along the ribbon and $N_y = 300$ perpendicular to the ribbon. In order to limit the numerical effort, we limit ourselves to the double Ising wall limit and with a fixed ratio $x_0 = 0.1 w$ of wall parameters. For the choice $w=0.2$, we have the same parameters as in section \ref{subsec:1dDoubleIsingWall} that allows a direct comparison to Fig.~\ref{fig6}(c) and (d).

In order to connect with the result of the previous section, we start with the boundary conditions (ne) of Fig.~\ref{figBC} corresponding to no environment. We concentrate on the two eigenstates with the lowest positive eigenvalues, $\varepsilon_1$ and $\varepsilon_2$, whose dependence on the parameters $h_0$ and $h_{\rm ex}$ are shown in Fig.~\ref{fig:Num1}. These eigenenergies are basically zero in the regime $|h_0 - h_{\rm ex}| > h_{cr}$ where the superconductor is gapless. In the inner parameter regime $|h_0 - h_{\rm ex}| < h_{cr}$, the superconductor is instead gapped and states localized to the domain wall arise. 
The bound state arising at small fields $h_0 = h_{\rm ex}$, on the one hand,  has a finite energy $\varepsilon_1$. This bound state supports excitations that propagate along the domain wall, the lowest of which is represented by $\varepsilon_2$. This excitation only costs negligible energy such that $\varepsilon_2 - \varepsilon_1 \approx 0$.
On the other hand, the bound state materializing at higher fields $h_0 = h_{\rm ex}$ possesses zero energy $\varepsilon_1 \approx 0$. As we will demonstrate below, this corresponds to the energy of a zero Majorana mode that is localized at the edges of the ribbon. The energy difference $\varepsilon_2 - \varepsilon_1$ corresponds to the topological gap which should be compared to the values within this region shown in Fig.~\ref{fig6}(d).

The study of a finite-size system allows in particular to study the spatial extent of the Majorana wavefunction. For this purpose, we consider the lowest energy eigenfunction $\vec \phi(\vec r)$, and we discuss the corresponding 
local density $\rho(\vec r) = \vec \phi^\dagger(\vec r) \vec \phi(\vec r)$ as well as its Majorana polarization $P(\vec r) = \vec \phi^\dagger(\vec r) \mathcal{C} \vec \phi(\vec r)$ introduced by Sedlmayr~et~al.\cite{Sedlmayr:2015cza}, where $\mathcal{C} = \tau_y \sigma_y \mathcal{K}$ is the particle-hole operator where $\mathcal{K}$ implements complex conjugation. Note that $P$ is complex as $\mathcal{C}$ is an anti-unitary operator. As a first indicator, we show in Fig.~\ref{fig:Num1}(d) 
the dependence of the integrated absolute value of the polarization $\int d \vec r |P(\vec r)|$ on the parameters $h_0$ and $h_{\rm ex}$.
This value is expected to be unity for a localized Majorana mode \cite{Sedlmayr:2015cza}. The results in Fig.~\ref{fig:Num1}(d) corroborate the appearance of MBSs within the previously identified region.

\begin{figure}
    \centering
    \includegraphics[width=0.9\linewidth]{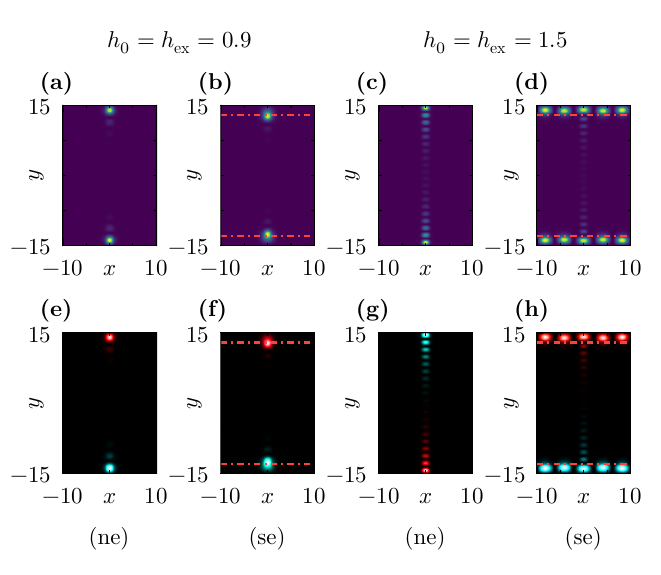}
    \caption{Local density $\rho(\vec r) = \vec \phi^\dagger(\vec r) \vec \phi(\vec r)$ (a) - (d) and Majorana polarization $P(\vec r) = \vec \phi^\dagger(\vec r) \mathcal{C} \vec \phi(\vec r)$ (e)-(h) for the lowest energy eigenstate $\vec \phi$ 
for two points within the MBS region of the phase diagram: (a), (b), (e) and (f) $h_0 = h_{\rm ex} = 0.9$ and  (c), (d), (g) and (h) $h_0 = h_{\rm ex} = 1.5$. For both sets of values, a MBS exist for (ne) boundary conditions in panel (a), (c), (e) and (g) with well localized density and Majorana polarization. For (se) boundary conditions, the Majorana exists in panel (b) and (f) but it is spilling into the superconducting environment (se) in panel (d) and (h).  
The Majorana polarization $P$ in the lower row is plotted as a complex density plot where the brightness and color corresponds to the magnitude $|P|$ and phase of $P$, respectively. The red and blue color thus indicate the different sign of $P$ for the Majoranas localized on different edges. Here, the domain width parameter $w = 0.8$.
}\label{fig:Num2}
\end{figure}

In Fig.~\ref{fig:Num2} we analyze $\rho(\vec r)$ and $P(\vec r)$ for values of $h_0$ and $h_{\rm ex}$ well placed within the expected MBS regime in more detail. 
We use here a slightly larger wall parameter $w = 0.8$ where the MBS regime materializes already for smaller values of $h_0 = h_{\rm ex}$ and the impact of boundary conditions is more pronounced.
Fig.~\ref{fig:Num2}(a) and (c) demonstrate that the local density of the eigenstate with the lowest energy is indeed well localized at the edges of the domain wall positioned at $x=0$ and extending along the $y$ direction. 
In addition, the eigenenergy is close to zero $\varepsilon_1 \sim 10^{-5}\Delta$, which can be even further decreased by increasing the width of the ribbon in the $y$-direction. Panel (e) and (g) of Fig.~\ref{fig:Num2} show the corresponding Majorana polarization where the brightness represents the magnitude $|P(\vec r)|$ and the color indicates the phase of $P(\vec r)$. The distinct color at the upper and lower edge signals the different signs of $P$ for the two corresponding Majorana states. These results were obtained for the (ne) boundary conditions of Fig.~\ref{figBC}. The boundary condition (se) consisting of a superconducting environment is expected to disfavour MBSs according to the necessary conditions summarized in Fig.~\ref{figNCOND}(b). The parameter values of panel Fig.~\ref{fig:Num2}(b) and (f) are located within the regime that fulfils the necessary conditions and indeed the Majorana state is well localized. In contrast, panel (d) and (h) are computed for values outside this regime and the Majorana is seen to be spilling into the environment.

We can summarize that the numerics on the finite size sample confirm the phase diagram in Fig.~\ref{fig6}(c) and (d) indicating that the red solid line indeed corresponds to a transition between a topologically trivial and non-trivial quasi-one dimensional superconducting wire where MBSs materialize at the edges of the ribbon.

 \begin{figure}
    \centering
    \includegraphics[width=\linewidth]{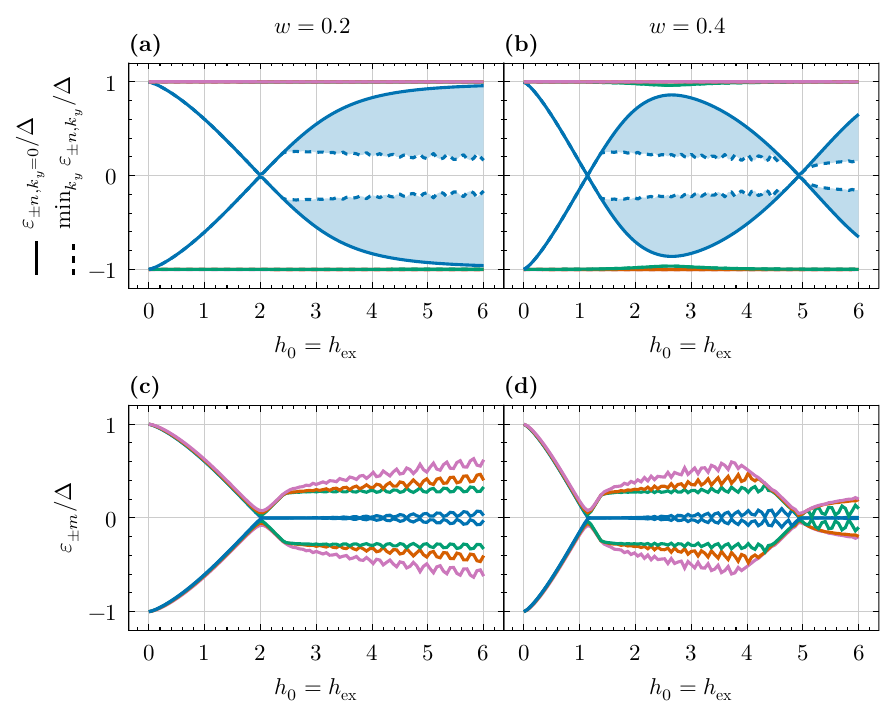}
    \caption{
 In-gap energy spectrum as a function of $h_0$ with $h_0 = h_{\rm ex}$ for two values of the width of the Ising double wall: (a), (c) $w = 0.2$ and (b), (d) $w = 0.4$. Panels (a) and (b) display the eigenenergies $\varepsilon_{\pm,n,k_y}$ of the Fourier transformed Hamiltonian \eqref{eq:hamFT} at wavevector $k_y = 0$ (solid lines) as well as its minimum value (dashed lines). Panels (c) and (d) show the lowest eight eigenenergies $\varepsilon_{\pm,m}$ obtained from the full numerical diagonalization of the finite system. Note that in the latter case, there exist many more in-gap states corresponding to modes propagating with finite $k_y$ along the domain wall. The wiggles are attributed to a finite size effect.}     
  \label{fig:width}
\end{figure}

Finally, we connect with the discussion in section \ref{subsec:1d2PiWall} and address the modifications to the phase diagram when leaving the double Ising wall limit and approaching the $2\pi$ domain wall regime. We compare two domain wall widths $w = 0.2$ and $w = 0.4$ with fixed ratio $x_0/w = 0.1$. As the lattice constant $a = 0.1$, both cases are still in the double Ising wall limit, $a \gg x_0$ but illustrate the tendency when approaching the $2\pi$ domain wall regime. 

Fig.~\ref{fig:width}(a) and (b) show the spectrum $\varepsilon_{\pm,n,k_y}$ obtained from the partially Fourier transformed Hamiltonian \eqref{eq:hamFT}: the solid line shows the value at $k_y = 0$ and the dashed line represents the minimal value  min$_{k_y} |\varepsilon_{\pm,n,k_y}|$. The vanishing of the eigenenergy in panel Fig.~\ref{fig:width}(a) at $h_0 \approx 2.0$ for $w = 0.2$ is consistent with the red solid line in Fig.~\ref{fig6}(c). This critical field shifts to lower values $h_0 \approx 1.2$ for $w = 0.4$, see Fig.~\ref{fig:width}(b). Moreover, the  eigenenergy vanishes again at a second critical field at $h_0 \approx 4.4$,  implying that the topological phase housing the MBSs is only stable over a finite field range. Note that the eigenenergy vanishes for a wavevector $k_y = 0$ although well within the MBS phase the spectrum $\varepsilon_{+,1,k_y}$ becomes minimal at a finite $k_y \neq 0$.
In order to elucidate the state emerging at larger fields, we show in Fig.~\ref{fig:width}(c) and (d) the lowest eight eigenenergies obtained from the full numerical diagonalization of the finite system. The wiggly oscillations are a finite size effect that also hybridizes the Majorana modes particularly for larger field values. When crossing the first critical field, a single pair of zero energy states materializes, indicating the presence of MBSs. 
The kink at $h_0 \approx 2.4$ in Fig.~\ref{fig:width}(c) indicates that a finite wavevector $k_y$ emerges where the spectrum obeys $\varepsilon_{+,1,k_y} < \varepsilon_{+,1,k_y = 0}$, thus yielding smaller values than the solid line in Fig.~\ref{fig:width}(a). The second kink at $h_0 \approx 4.5$ in Fig.~\ref{fig:width}(d) suggests that $\varepsilon_{+,1,k_y = 0}$ becomes minimal again reproducing the same value for the second critical field as in Fig.~\ref{fig:width}(b). After crossing the second critical field, an additional pair of states with energy close to zero is found suggesting the presence of further edge states, see Fig.~\ref{fig:width}(d). We anticipate however that these pairs of edge states are unstable with respect to perturbations, as the topology of the system eventually belongs to the class D. 
 
\section{Summary and Discussion}
\label{sec:conclusion}

\begin{figure}[t]
    \includegraphics[width=0.48\textwidth]{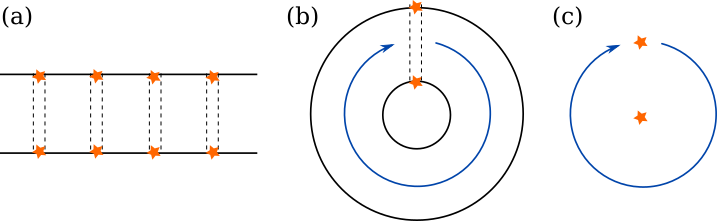}
    \caption{\label{fig:perspectives} (a) A sequence of topological $2\pi$ domain walls (dashed lines) in a magnetic ribbon on a superconductor will host a chain of MBS (orange stars) at each rim of the ribbon, resembling Kitaev chains. (b) If the ribbon takes the shape of an annulus, moving a domain wall around the device is topologically equivalent to (c), the braiding of the inner and outer MBS.}
\end{figure}

Magnetic-superconducting heterostructures provide platforms for the realization of Majorana zero modes. In this work, we considered a magnetic ribbon that is coupled via an antiferromagnetic exchange interaction to a two-dimensional superconductor with Rashba spin-orbit coupling. In order to minimize orbital pair breaking effects, we assumed that the magnetization is forced to be in-plane due to the presence of a strong easy-plane anisotropy. An interesting situation arises if the superconductor is in a topologically trivial 
phase for the fully polarized ground state of the ferromagnet. In this case, a $2\pi$ domain wall is able to induce an in-gap state within the superconductor that is localized on the domain wall. This realizes a quasi-one dimensional superconducting wire. As a function of an applied magnetic field, a phase transition can be induced where this quasi-1d wire converts between a topologically trivial and non-trivial state. The former will host zero Majorana edge modes that are localized at the terminations of the wire, i.e., at the end points of the domain wall at the edges of the ribbon, see Fig.~\ref{fig:intro}. 

We investigated the robustness and the feasibility of this proposal for a set of representative parameters specified in section \ref{Subsec:Parameters}. The main result is summarized in Fig.~\ref{fig6}(c): for a given antiferromagnetic exchange field $h_{\rm ex}$ a phase with Majorana bound states (MBS) can be achieved by tuning the externally applied field $h_0$ up to a field that is on the order of 
$h_{\rm cr}$ defined in Eq.~\eqref{eq:hcr} that, in appropriate units, is itself on the order of the superconducting gap. The larger $h_{\rm ex}$ the larger is the extent of the MBS phase as a function of $h_0$. Whereas most of our numerical computations in fact considered a sharp double Ising wall with a width on the order of the atomic lattice spacing $a$, we also demonstrated explicitly in Fig.~\ref{fig1dWire2PiWall} that this scenario is robust when considering a proper $2\pi$ domain wall whose width is much larger than and incommensurate with $a$.

Majorana bound states localized at the edges of an incommensurate $2\pi$ domain wall offer an interesting perspective for manipulating Majorana zero modes. Such incommensurate $2\pi$ domain walls spontaneously break an effective translational invariance in the transversal direction along the ribbon. As a result, they can be easily moved by applying spin currents as they are only weakly pinned by disorder. This has been demonstrated experimentally already more than ten years ago \cite{Ryu2013,Emori2013}. This is in contrast to Ising walls that are commensurate with the crystal lattice and, consequently,  are much stronger bound to the crystal potential. 

In order to conserve the coherence of the Majorana modes and thus enable Majorana operations, a $2\pi$ domain wall motion induced by spin currents needs to fulfil some requirements. The motion should be adiabatic in the sense that the kinetic energy generated by a finite domain wall motion velocity needs to be smaller than the characteristic energies of the system. First, it should be smaller than the topological gap in order to avoid mixing the Majorana modes with higher energy states. Moreover, the quasi-1d superconducting wire might host various bands and the kinetic energy should be also smaller than the band gap in order to avoid band-mixing effects. 

In any case, a major advantage of the system that we propose is the high level of control of domain walls with existing experimental techniques. The generation, placement, and motion of magnetic domain walls are all well-established routines in spintronics. In the following, we describe two perspective applications of our proposal to illustrate its flexibility: tunable Kitaev chains \cite{Kitaev:2000nmw} and a geometry suitable for real-space braiding of MBS.

First, let us consider a series of domain walls in a ribbon, see Fig.~\ref{fig:perspectives}(a). The domain walls can be moved by currents \cite{Ryu2013, Emori2013}, placed in an arbitrary sequence by a magnetic STM tip, or adjusted by an external magnetic field. For a chiral magnet in an external field, for instance, the spacing between adjacent domain walls can easily be tuned by the field strength. Thereby, the distance between the MBS at each edge of the ribbon can be varied. For small distances, the overlap of Majorana wave functions will cause a hybridization into one-dimensional modes. The advantages of this type of bottom-up construction of a Kitaev chain from actual MBS were recently discussed in~\cite{Marra:2021xqk, Marra:2021nrw}. Here, manipulating the domain wall positions would allow adjusting the coupling of Majorana modes and place the chain as a whole on either side of the topological transition.

As a second example, consider a circular geometry with a magnetic ribbon shaped as an annulus on the superconducting substrate, see Fig.~\ref{fig:perspectives} (b)-(c). Moving a topological domain wall once around the system is then topologically equivalent to the outer MBS traveling once around the inner MBS. Hence, our proposal allows in principle for braiding operations.

\begin{acknowledgments}
This work was funded by the Deutsche Forschungsgemeinschaft (DFG, German Research Foundation) -- 445312953 (S.R.~and M.G.); 429691603 (S.R.); and 403030645 (M.G.). The authors also acknowledge support by the state of Baden-Württemberg through bwHPC.
\end{acknowledgments}

\bibliography{literature}

\end{document}